\documentclass[a4paper,12pt]{article}
\usepackage[cp1251]{inputenc}
\usepackage[dvips]{graphicx}
\graphicspath{{.}}

\title{Orientational effect of uniaxially deformed aerogel on the order parameter of superfluid~$^3He$}
\author {E.V. Surovtsev \\
{\it P.L.Kapitza Institute for Physical Problems,}
\\ {\it ul. Kosygina, 2, Moscow 119334, Russia}}
\date{ }

\begin{document}
\maketitle

\mathsurround=2pt \sloppy
\begin{abstract}
Effect of uniaxially-deformed aerogel on orientation of the order
parameter of superfluid $^3He$ is considered. Approach used
\cite{F1} takes into account correlations  in positions of
particles forming aerogel. Calculations show that for the case of
the B-like phase in the uniaxially-stretched aerogel deformation
on the order of $10^{-2}\div10^{-4}$ is already strong enough to
achieve orientation of the order parameter which is different from
that required by magnetic field in bulk liquid. For the A-like
phase in uniaxially-squeezed aerogel the order of magnitude of the
orientational effect is estimated. It is shown that the
correlations in positions of particles forming aerogel has
stronger effect on the orientation of the order parameter than on
the transition temperature.
\end{abstract}

Superfluid $^3He$ is a convenient system for investigation of the
effect of impurities on formation and orientation of the order
parameter. Interaction between aerogel and the order parameter of
$^3He$ can be revealed in suppression of the superfluid transition
temperature, as well as in possible changing of a form of the
order parameter and spatial orientation of the order parameter.
The first attempts to evaluate the effect of the anisotropy of
aerogel on the orientation of the order parameter were based on
the Homogeneous Scattering Model (HSM) Ref.\cite{SF1}, which is a
generalization of the Abrikosov-Gor'kov theory of superconducting
alloys for the case of p-wave pairing. But this theory does not
take into account correlations in positions of particles forming
aerogel. Recently I.A. Fomin has shown, that such correlations can
play an important role effecting on the
transition temperature to the superfluid state Ref.\cite{F1}. 
In the present paper the effect of uniaxially deformed aerogel on
the orientation of the order parameter is considered within the
same approach.

In what follows we restrict our consideration to a temperature
region in the vicinity of the bulk transition temperature $T_b$.
In this region for our purposes one can use linearized Ginzburg
and Landau equation. Linearized Ginzburg and Landau equation in
momentum representation in the notations of Ref.\cite{F1} can be
written in the following form:
\begin{eqnarray}
\label{GL} \left( \tau \delta_{jl}-\frac{3}{5}\xi_s^2
k^2(\delta_{jl}+2k_j k_l)\right)a_{\mu l}(\mathbf{k})=\int a_{\mu
l}(\mathbf{q})
\eta_{jl}(\mathbf{k-q})\frac{d^3 q}{(2\pi)^3},\\
A_{\mu j}(\mathbf{r})=\int a_{\mu j}(\mathbf{q}) \exp(-i
\mathbf{q}\mathbf{r})\frac{d^3 q}{(2\pi)^3},~~
\eta_{jl}(\mathbf{r})=\int \eta_{jl}(\mathbf{q})\frac{d^3
q}{(2\pi)^3} \nonumber
\end{eqnarray}
where $\tau=\frac{T-T_b}{T_b}$, $A_{\mu j}$ - $3\times3$ complex
matrix, corresponding to the order parameter in the case of p-wave
pairing, $\eta_{jl}(\mathbf{r})$ - real symmetric tensor, which
characterizes interaction between impurities and the order
parameter. The form of the tensor $\eta_{jl}(\mathbf{r})$ depends
on a structure of aerogel and a type of quasiparticle scattering
by aerogel particles. We assume here, that aerogel consists of
spheres with uniform radii $\rho$, distributed with an averaged
density$~$$n$. For simplicity the type of scattering is chosen to
be diffusive. Assuming low concentration of impurities and using
the Rainer and Vuorio theory of small objects in superfluid $^3He$
one can find the tensor $\eta_{jl}(\mathbf{r})$ to be:
\begin{equation} \eta_{jl}(\mathbf{r})=\sum_s
\eta_{jl}^1(\mathbf{r}-\mathbf{r_s}),
\end{equation}
where $s$ - number of impurity,
\begin{equation}
\eta_{jl}^1(\mathbf{r})=-\frac{\rho^2}{r^2}\nu_j\nu_l \ln \left[
\tanh \left( \frac{r}{2\xi_0}\right) \right].
\end{equation}
As it was shown in Ref.\cite{F1}, for the natural parameters of
aerogel and superfluid $^3He$ the rhs of (\ref{GL}) can be treated
as perturbation.

In order to find orientational correction to the free-energy one
has to solve the problem of eigenvalues of Eq. (\ref{GL}). In the
absence of perturbation the eigenvalues of Eq. (\ref{GL}) are
degenerate, i.e single temperature transition corresponds to
different orbital components of the order parameter (different
orientations of the order parameter). Nominally isotropic aerogel
produces isotropic perturbation. This situation was considered in
Ref.\cite{F1}.   
In the case of such isotropic perturbation the eigenvalues of
Ginzburg and Landau equation remain degenerate. In the presence of
anisotropic perturbation induced by uniaxially deformed aerogel
this degeneracy is partly lifted. Solution of the secular equation
of perturbation theory yields corrections to the transition
temperature, which correspond to "longitudinal"$~$(along the
direction of deformation) and "transverse"$~$transition
temperatures. Therefore a term with a tensor of transition
temperature appears in the free-energy. The diagonal terms of this
tensor are given by solution of the secular equation. After
separation of isotropic and anisotropic contributions to the
transition temperature orientational term in the free-energy can
be expressed in a form:
\begin{equation}
\label{free_energy}
\tau_{jl}^a A_{\mu j}A_{\mu l}^{*},
\end{equation}
$\tau^a_{ll}=0.$ Below the method of Green's functions is used.

\begin{figure}
\begin{center}
\includegraphics[%
  width=0.6\linewidth,
  keepaspectratio]{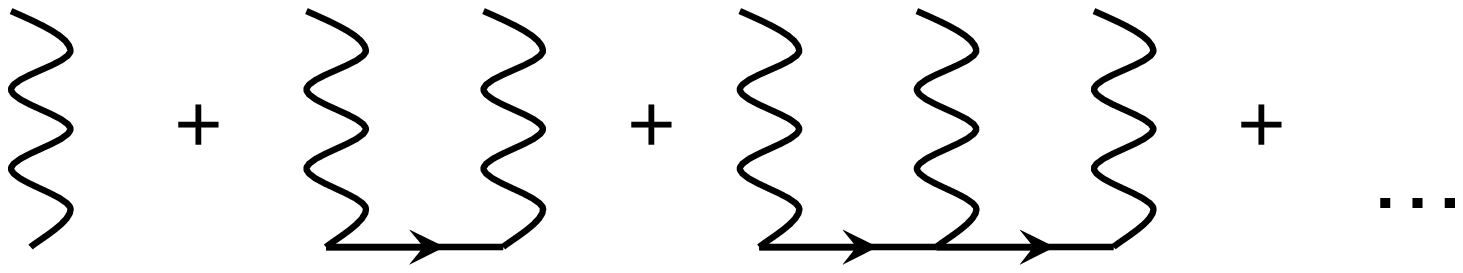} 
\end{center}
\caption ~
\label{frame1}
\end{figure}

The Green's function tensor of Eq. (\ref{GL}), averaged over all
realizations of the potential $\eta_{jl}(\mathbf{r})$, has the
form:
\begin{eqnarray}
\label{GreenF} \langle
G_{jl}(\tau,\mathbf{k},\mathbf{k}^{'})\rangle=(2\pi)^3\delta(\mathbf{k}-\mathbf{k}^{'})
\left[\left(G_{jl}^{(0)}(\tau,\mathbf{k})\right)^{-1}-\langle\Sigma_{jl}(\tau,\mathbf{k})\rangle\right]^{-1},
\end{eqnarray}
where
\begin{equation}
\left(G_{jl}^{(0)}(\tau,\mathbf{k})\right)^{-1}=\tau
\delta_{jl}-\frac{3}{5}\xi_s^2 k^2(\delta_{jl}+2k_j k_l),
\end{equation}
is the unperturbed Green's function tensor and the tensor of the
Self-energy part is shown diagrammatically on the
Fig.\ref{frame1}. Arrows on the figure represent the unperturbed
Green's function tensor, and wavy lines correspond to the impurity
potential:
\begin{equation}
 \eta_{jl}(\mathbf{k}-\mathbf{k^{'}})=
\eta_{jl}^1(\mathbf{k}-\mathbf{k^{'}})\sum_s
e^{i(\mathbf{k}-\mathbf{k^{'}})\mathbf{r_s}}.
\end{equation}
Integration is made over momenta of internal lines. It is
significant that the expression (\ref{GreenF}) differs from the
corresponding expression of Ref.\cite{F1} in the way that it is
not proportional to $\delta_{jl}$, but it has more complicated
tensor structure.

The eigenvalues of Eq. (\ref{GL}) are determined by the poles of
Green's function, which are found from the secular equation:
\begin{equation}
\det\left( \delta_{jl}\cdot\tau-\frac{3}{5}\xi_s^2
k^2(\delta_{jl}+2k_j
k_l)-\Sigma_{jl}(\mathbf{k})\right)_{\mathbf{k}\rightarrow 0}=0.
\end{equation}
Here the limit $k\rightarrow0$ corresponds to the "mobility edge"
 Ref. \cite{F1}.

The first order correction to the Self-energy part comes from the
first term of series Fig.\ref{frame1}. Averaging over coordinates
of particles renders:
\begin{equation}
\tau^{(1)}_{ba}= \frac{1}{3} \eta^{(1)}_{ll}(\mathbf{k}\rightarrow
0) n=\frac{\pi^2}{4}\frac{\xi_0}{l_{tr}},
\end{equation}
where $l_{tr}$ - mean free path, $\xi_0$ - coherence length of
superfluid $^3He$.

The effect that we are interested in appears starting from the
second order correction:
\begin{equation}
\label{second} \Sigma_{jl}^{(2)}(\mathbf{k})=\int
\eta^{(1)}_{jm}(\mathbf{k}-\mathbf{k_1})\eta^{(1)}_{ln}(\mathbf{k_1}-\mathbf{k})n\langle\sum_t
e^{i(\mathbf{k_1}-\mathbf{k})\mathbf{r_{st}}}\rangle
G^{(0)}_{mn}(\tau=0,\mathbf{k_1})\frac{d^3 k_1}{(2\pi)^3},
\end{equation}
where averaged sum
\begin{equation}
\langle \sum_{t}
e^{i(\mathbf{k_1}-\mathbf{k})\mathbf{r_{st}}}\rangle=S(\mathbf{k_1}-\mathbf{k}),
\end{equation}
equals by definition to the structure factor of aerogel, which
characterizes correlations in positions of particles forming
aerogel. The structure factor can be expressed in terms of the
pair correlation function via Fourier transformation:
\begin{equation}
S(\mathbf{k})=n\int C(\mathbf{r})e^{i\mathbf{k}\mathbf{r}}d^3r.
\end{equation}
In accordance with the general rules of perturbation theory one
substitutes in (\ref{second}) the unperturbed Green's function,
what is made by implying the condition $\tau=0$.

Deformation of aerogel changes its structure factor. It is assumed
here that non-deformed aerogel is isotropic. As a result its
structure factor does not depend on the direction of wave-vector
$\mathbf{k}$. Aerogel has several typical lengthscales. On the
scales greater than several thousands ${\AA}$ aerogel is
homogeneous. There is also an interval of scales where aerogel
reveals its fractal structure. It means, that in the corresponding
interval of distances $\rho<r<R$ its pair correlation function has
fractal behavior:
\begin{equation}
\label{Cor_func1} C(\mathbf{r})=A(R/r)^{3-D},
\end{equation}
where D - the fractal dimension of aerogel, $R$ - the correlation
radius of aerogel, i.e. distance whereon correlation between
particles decreases. In order to provide smooth decreasing of
correlation function on the distances of correlation radius
$C(\mathbf{r})$ is replaced by the following model correlation
function:
\begin{equation}
\label{Cor_func2} C(\mathbf{r})\longrightarrow
C(\mathbf{r})\exp(-r/R).
\end{equation}
\begin{figure}
\begin{center}
\includegraphics[%
  width=0.55\linewidth,
  keepaspectratio]{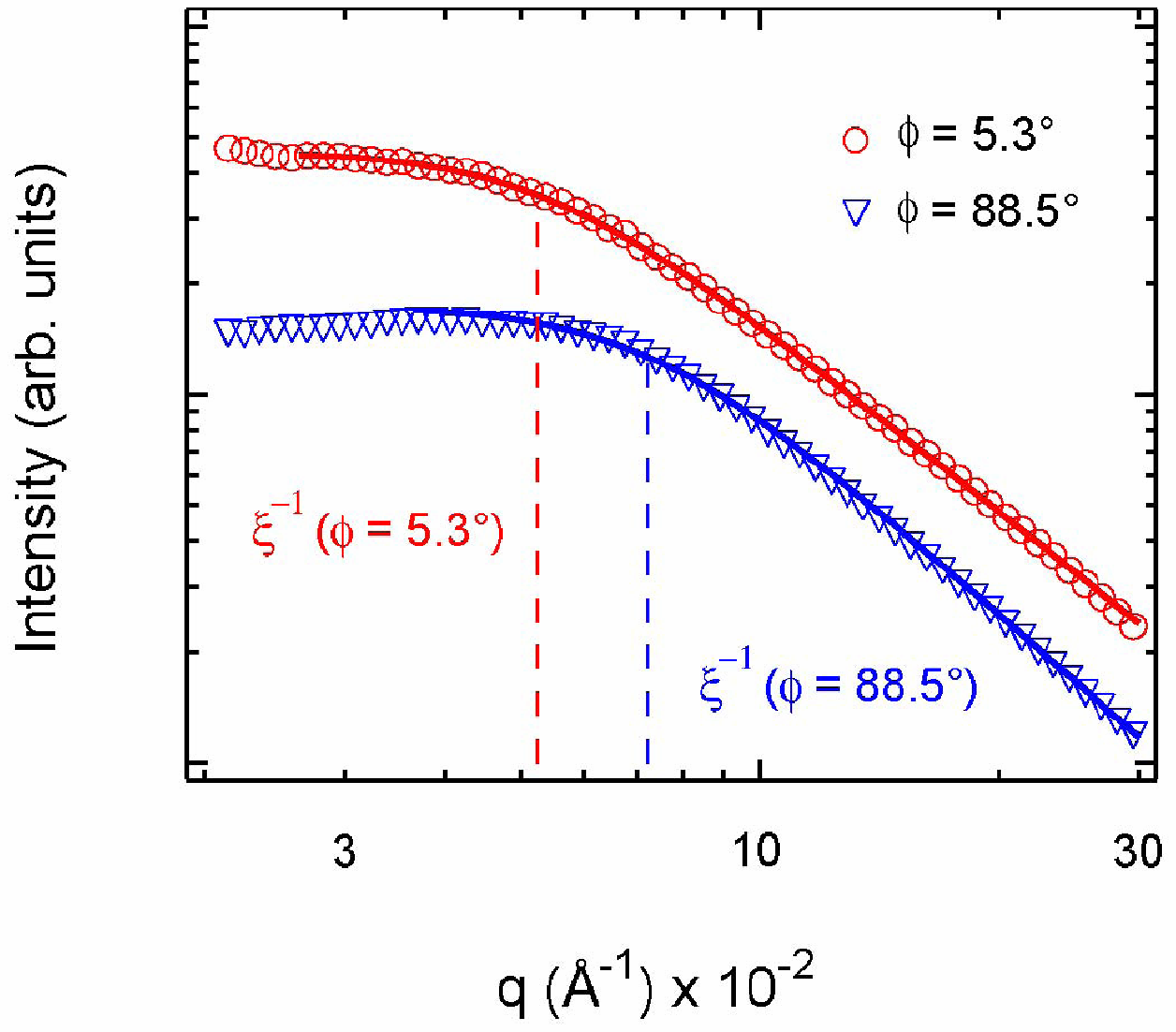} 
\end{center}
\caption ~ Dependence of the X-ray scattered intensity
$I(q,\phi=\theta+\frac{\pi}{2})$ on the momentum transfer, $q$,
for two values of the angle $\theta$. The curves have been offset
vertically for clarity, otherwise the data points would coincide
at high $q$. Ref.\cite{JP}
 \label{frame2}
\end{figure}
The coefficient $A$ in (\ref{Cor_func1}), (\ref{Cor_func2}) is
found from normalization condition Ref.\cite{F1}. It is
significant that the structure factor can be directly measured in
the small angle X-ray scattering measurements. The results of such
measurements for an uniaxially deformed aerogel are represented in
Ref.\cite{JP}. As it was shown by the authors of Ref.\cite{JP} the
results of their measurements can be fitted if the assumption is
made that the fractal dimension of aerogel is not changed by the
deformation and the correlation radius gains the following angular
dependence (fig.\ref{frame2}):
\begin{equation}
R(\theta)=R_0+r_1\cos(2\theta),
\end{equation}
where $R_0$ - the correlation radius before deformation, $\theta$
- the angle between deformation axis and chosen direction.
Amplitude of changing of correlation radius can be expressed via
macroscopic parameters characterizing strain of aerogel:
\begin{equation}
\frac{r_1}{R_0}=\frac{1}{2}(1-\nu)\alpha\frac{\Delta l}{l_0},
\end{equation}
here $\frac{\Delta l}{l_0}$ relative change of aerogel length,
$\nu$ - the macroscopic Poisson ratio, ($\nu\approx 0.2\div 0.3$),
$\alpha$ - the coefficient of transmission of macroscopic strain
down to the level of correlation length. It was found that
$\alpha$ approximately equals to 3.

Substitution of the new structure factor into (\ref{second})
yields the anisotropic correction to the Self-energy part. Its
principal order contribution arises from angular dependence of the
correlation radius and k-dependence of the unperturbed Green's
function. For the anisotropic part of the Self-energy function in
the second order correction one arrives at:
$$
\Sigma_{jl}^{(2)a}\approx
0.35\left(\frac{\pi}{3}\right)^4\frac{3-D}{D-1}\frac{R_0\cdot
r_1}{l^2_{tr}}\left(
\begin{array}{ccc}
1&0&0\\
0&1&0\\
0&0&-2
\end{array}
\right). \eqno \stepcounter{equation}(\arabic{equation})
$$
The third order correction has the form:
\begin{equation}
\Sigma_{jl}^{(3)}=n\left(\eta^{(1)}(0)\right)^3\int\frac{d^3k_1}{(2\pi)^3}\frac{d^3k_2}{(2\pi)^3}G^{(0)}_{jm}(\mathbf{k_1})G^{(0)}_{lm}(\mathbf{k_2})
\langle\sum_{t, u}e^{i \mathbf{k_1}\mathbf{r}_{st}}e^{i
\mathbf{k_2}\mathbf{r}_{su}}\rangle
\end{equation}
According to Ref.\cite{F1} it can be calculated with the
assumption that:
\begin{equation}
\langle\sum_{t,u}e^{i\mathbf{k_1}\mathbf{r_{st}}}e^{i\mathbf{k_2}\mathbf{r_{us}}}\rangle=S(\mathbf{k_1})S(\mathbf{k_2}).
\end{equation}
If the same assumption is made for the high order corrections, sum
of series, which forms geometric progression, can be found. In
particular for the anisotropic part of the Self-energy one has:
$$
 \tau^a_{jl}=\Sigma_{jl}^{a}\approx
0.35\left(\frac{\pi}{3}\right)^4\frac{3-D}{D-1}\frac{R\cdot
r_1}{l^2_{tr}}\displaystyle\frac{1}{\left(1+\frac{25}{27}\frac{\pi^2}{2}\frac{R^2}{\xi_0
l_{tr}}\frac{1}{D-1}\right)^2}\left(
\begin{array}{ccc}
1&0&0\\
0&1&0\\
0&0&-2
\end{array}
\right). \eqno \stepcounter{equation}(\arabic{equation})
\label{answer}
$$
As it is seen from the final formula in the case when $\xi_0\gg R$
HSM can be used. In the opposite case correlations play an
important role in the orientational effect.

Let us first consider  the orientational effect of the anisotropic
aerogel on the order parameter of the B-like phase. In the absence
of magnetic field the order parameter of the B-like phase is
isotropic and no orientational effect is expected. But situation
changes in the presence of a magnetic field. The order parameter
of the B-like phase in magnetic field has the form \cite{WV}:
\begin{equation}
\label{order_parameter}
A_{\mu
j}=\frac{e^{i\phi}}{\sqrt{3}}\left(\Delta_{\bot}R_{\mu
j}+(\Delta_{\|}-\Delta_{\bot})R_{\nu
j}\hat{h}_{\nu}\hat{h}_\mu\right),
\end{equation}
the rotation matrix $R_{\mu j}(\mathbf{\hat{n}},\theta)$ is
characterized by a rotation axis $\mathbf{\hat{n}}$  and the
rotation angle
$\theta_L=\arccos(-\frac{1}{4}\frac{\Delta_{\parallel}}{\Delta_{\perp}})$,
which is found from minimization of dipole energy,
$\mathbf{\hat{h}}$ is the direction of magnetic field. Orientation
of $\mathbf{\hat{n}}$ is induced by magnetic field and dipole
forces. The corresponding term in the free energy equals to:
\begin{equation}
\Delta F_H^B=\lambda_D
N_F\left(\Delta_{\|}-\Delta_{\bot}\right)\Delta_{\bot}(\hat{n}\cdot\hat{h})^2.
\end{equation}
At the same time globally anisotropic aerogel induces orientation
of the orbital part of the order parameter. Substitution of
(\ref{order_parameter}) into (\ref{free_energy}) gives:
\begin{equation}
\label{Orient_Energy_B} \Delta F_{a}^B=
\frac{1}{3}N_F\left(\Delta^2_{\|}-\Delta^2_{\bot}\right)R_{\mu
j}R_{\nu l}h_\mu
h_{\nu}\tau^{a}_{jl}\approx\frac{2}{3}\left(\Delta_{\|}-\Delta_{\bot}\right)\Delta_{\bot}\Sigma^a_{xx}(1-3l_z^2),
\end{equation}
here the orbital vector
\begin{equation}
\label{l} l_j=\hat{R}_{\mu j}S_{\mu},
\end{equation}
is introduced, $S_\mu$ - the spin vector.
\begin{figure}
\begin{center}
\includegraphics[%
  width=0.75\linewidth,
  keepaspectratio]{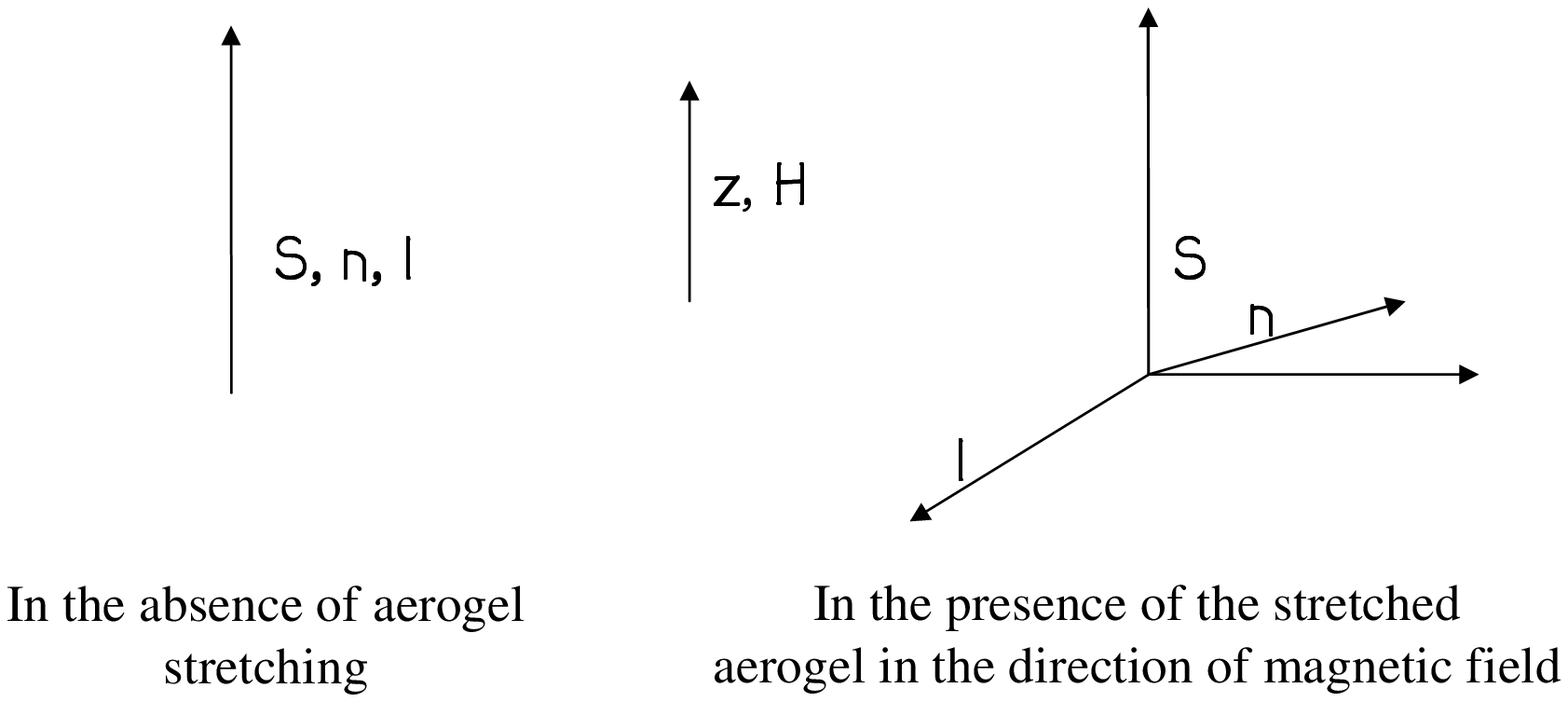}
\end{center}
\caption ~ Changing in orientation of vectors $\mathbf{\hat{n}}$
and $\mathbf{\hat{l}}$ in the presence of uniaxially-stretched
aerogel. \label{frame3}
\end{figure}
Let us compare orientational energies for the case of uniaxially
stretched aerogel in the direction of the magnetic field.  In the
absence of anisotropy vectors $\mathbf{\hat{n}}$
$\mathbf{\hat{l}}$ and $\mathbf{\hat{s}}$ are directed along the
magnetic field fig.\ref{frame3}. When aerogel is stretched vector
$\mathbf{\hat{l}}$ tends to turn in the direction perpendicular to
the magnetic field. To realize this it is needed to turn vector
$\mathbf{\hat{n}}$. Taking into account the minimum of dipole
energy solutions of Eq. (\ref{l}) for a given configuration of
vectors $\mathbf{\hat{n}}$ $\mathbf{\hat{l}}$ and
$\mathbf{\hat{s}}$ always exist. The new configuration starts to
form when:
\begin{equation}
\Delta F^B_{a}\sim\Delta F_H^B.
\end{equation}
Estimations made for the whole interval of parameters
characterizing aerogel and superfluid $^3He$, $D=1.6\div1.9$,
$R\sim200\div1400$ ${\AA}$,$\xi_0\sim170\div300$ ${\AA}$,
 $l_{tr}\approx1400$ ${\AA}$ show that deformation
on the order of
\begin{equation}
 \gamma_{min}\sim
5\cdot(10^{-3}\div 10^{-4})
\end{equation}
 is already strong enough to
achieve orientation of the order parameter which is different from
that required by the magnetic field in bulk liquid.

Estimation of the strength of the orientational effect in the
A-like phase can be made in a similar way. For clarity we assume
that aerogel is uniaxially squeezed. As it was shown in
Ref.\cite{F3}, \cite{Volovik} the order parameter of the A-like
phase in the case of uniaxially-squeezed aerogel has the form of
the bulk A-phase. Consequently the order parameter of the A-like
phase can be written in the form:
\begin{equation}
A_{\mu j}=\frac{1}{\sqrt{2}}d_\mu(m_j+in_j),
\end{equation}
where vector $\mathbf{\hat{d}}$ corresponds to the spin part of
the order parameter, and vector
$\mathbf{l}=[\mathbf{m},\mathbf{n}]$ is its orbital part. In the
absence of squeezed aerogel magnetic field and dipole forces
orient vector $\mathbf{\hat{l}}$ perpendicular to the magnetic
field:
\begin{equation}
\Delta F_H^A=\frac{1}{2}\Delta
\chi(\mathbf{\hat{d}}\cdot\mathbf{H})^2,
\end{equation}
\begin{equation}
\Delta f_D=-\frac{2}{5}\lambda_D N_F
\Delta_0^2(\mathbf{\hat{d}}\cdot\mathbf{\hat{l}})^2.
\end{equation}
However, as in the case $^3He-B$ aerogel can orient the orbital
part of the order parameter:
\begin{equation}
\label{Orient_Energy} \Delta
F_a^A=-N_F\frac{\Delta^2}{2}\Sigma^a_{xx}(1-3l_z^2).
\end{equation}
Comparison of $\Delta F_a^A$ with $\Delta F_H^A$ can be used for
estimation of the strength of the orientational effect.
Calculations show that for the melting pressure magnetic field
corresponding to one percent squeezing approximately equals to
80G.

Comparison of Eqns. (\ref{Orient_Energy_B}), (\ref{Orient_Energy})
with the analogous formula from Ref.\cite{SF1}  shows that
correlations can significantly decrease effect of anisotropy (on
the order of 10). Let us note also that the effect of correlations
on orientation of the order parameter is stronger  than the
similar effect on the transition temperature.

I am grateful to I.A. Fomin for fruitful discussion. This work is
partly supported by RFBR grant (no. 07-02-00214), by Ministry of
Education and Science of Russian Federation, Russian
Science-Support Fund and Landau Scholarship (A.F.) from
Forschungszentrum J\"{u}lih, Germany

\end{document}